# Self-learning kinetic Monte Carlo model for arbitrary surface orientations

Andreas Latz, Lothar Brendel and Dietrich E. Wolf
Department of Physics and Center for Nanointegration Duisburg-Essen (CeNIDE), University of Duisburg-Essen, D-47057 Duisburg, Germany

## ABSTRACT

While the self-learning kinetic Monte Carlo (SLKMC) method enables the calculation of transition rates from a realistic potential, implementations of it were usually limited to one specific surface orientation. An example is the fcc (111) surface in Latz et al. 2012, J. Phys.: Condens. Matter 24, 485005. This work provides an extension by means of detecting the local orientation, and thus allows for the accurate simulation of arbitrarily shaped surfaces. We applied the model to the diffusion of Ag monolayer islands and voids on a Ag(111) and Ag(001) surface, as well as the relaxation of a three-dimensional spherical particle.

## INTRODUCTION

In kinetic Monte Carlo (KMC) simulations the challenging task is to build up a correct rate catalogue. The self-learning kinetic Monte Carlo (SLKMC) method provides a way to combine the accuracy of rates calculated from a realistic potential with the efficiency of a rate catalog [1, 2]. Each time a process occurs for the first time its transition rate is calculated and inserted into a database. Previously calculated rates can be reused by utilizing a pattern recognition scheme.

We recently developed a three-dimensional SLKMC model for simulations on a (111) oriented fcc substrate [2]. In the present study we first expand the model to allow simulations on (001) oriented fcc surfaces, too. As an example, we compare the diffusion of Ag monolayer islands and voids on a Ag(111) and Ag(001) substrate. The inclusion of (001) oriented surfaces enabled us to simulate systems with arbitrary orientations: For each hopping process, we determine a local orientation, which determines the surface orientation best suitable for the description of the local environment of the process. At choice are the eight possible {111} and six possible {001} surfaces. The model is tested by simulating the relaxation of a three-dimensional spherical particle.

## SIMULATION MODEL

The dynamical evolution of the systems is simulated by a sequence of thermally activated single atom hops, using a standard KMC algorithm [3]. The atoms are restricted to an fcc lattice and can hop to empty neighbor sites.

The hopping rates are given by the Arrhenius law

$$\nu = \nu_0 \exp\left(-\frac{E_\mathrm{a}}{k_\mathrm{B} T}\right), \tag{1}$$

where $v_0$ denotes the attempt frequency and $E_a$ is the activation energy, separating the initial and final configuration in the potential-energy surface (PES) of the system. $k_B$ is the Boltzmann constant and $T$ is the temperature. We assume a commonly used constant value of $v_0 = 10^{12}$ s$^{-1}$. To describe the interactions between atoms, a many-body tight binding potential is used, which was fitted to experimental values by Cleri and Rosato [4].

$E_a$ is strongly dependent on the local environment of the hopping process considered. Whenever a local environment for an allowed process appears for the first time, the corresponding activation energy is calculated. Once calculated, activation energies are stored and can be reused by utilizing our recently presented pattern recognition scheme [2].

In our model, the occupations of the first two neighbor shells of the initial as well as the final hopping site characterize a local environment, which we call a *configuration* in the remainder. The first shell consists of the nearest-neighbors of these two sites; further sites, up to fourth nearest-neighbors form the second shell. In figure 1, the neighbor shells for a nearest-neighbor hop in a (001) plane are shown. Three numbers are used to identify the corresponding process. The first number states the hopping direction. The second and third numbers are bitwise representations of the first and second shell occupancies. Symmetry operations are used to map equivalent hop directions onto a single one used for the $E_a$ calculation.

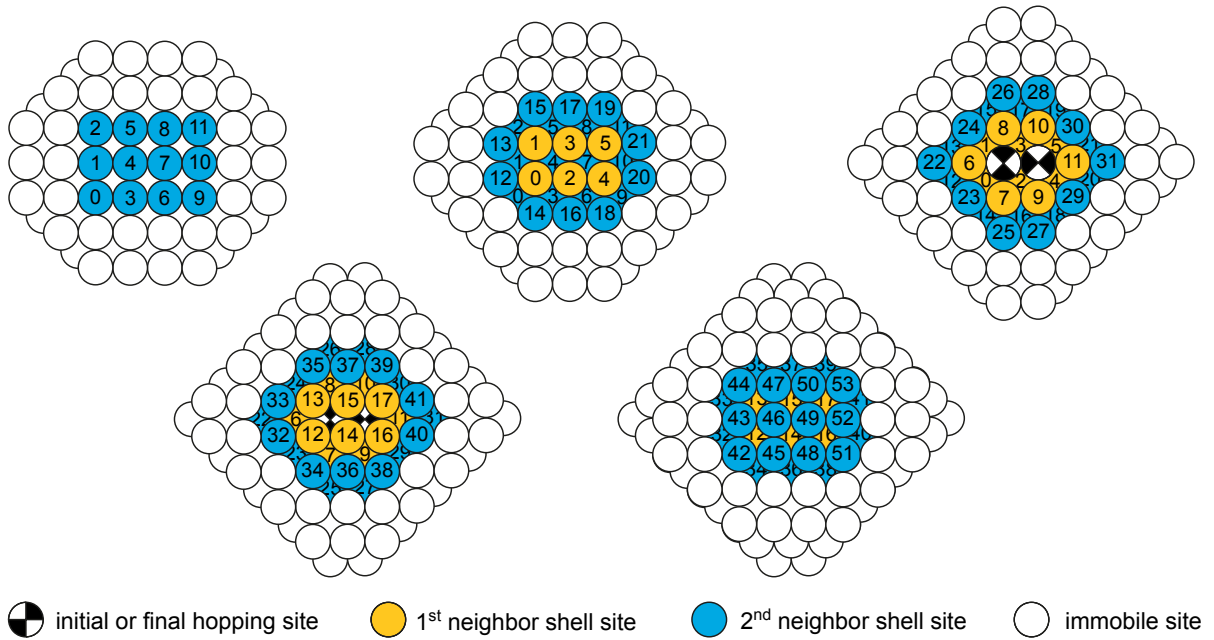

**Figure 1.** Local environment for nearest-neighbor hops in a (001) plane. Two neighbor shells around the initial and final hopping site are used to identify a configuration. The three-dimensional structure of the shells is made clear by stacking (001) layers as shown from the top left to the bottom right.

The activation energies are calculated using the drag method [5]. During the calculation, the first two neighbor shells are free to relax. To stabilize them, they need to be embedded in partially filled shells of immobile sites. These are filled in a way to represent a global surface orientation on which the process takes place. In our recent publication [2], we considered surfaces with an average (111) orientation. The modifications needed for (001) surfaces are nontrivial and will be explained next.

## (001) ORIENTED SURFACE

For the activation energy calculation we have to embed the first and second neighbor shells into the system's bulk. The embedding is constructed layer by layer, as described in [2], where these layers had the orientation (111). In general this leads to a configuration, which extrapolates the occupied sites of the first and second neighbor shell in the form of (111) terraces possibly separated by steps. The same algorithm can be applied using layers with (001) orientation. Of course, this leads to a different embedding, consisting of (001) terraces, as appropriate for surfaces, which globally have this orientation or a vicinal one. (Below we propose an algorithm that tells, which of the two embeddings is most appropriate locally for more complex surface morphologies.)

A difference to the (111) case is the existence of important third nearest-neighbor hops, which are not representable by two consecutive nearest-neighbor hops. Such a process is for example the detachment of an atom out of a (001) surface, shown in figure 2. The temporary nearest-neighbor position in between (red site) is neither a stable position nor a saddle point in the PES, and thus a corresponding transition rate cannot be calculated. Instead, the third nearest-neighbor processes linking the sites marked in black and white are now included.

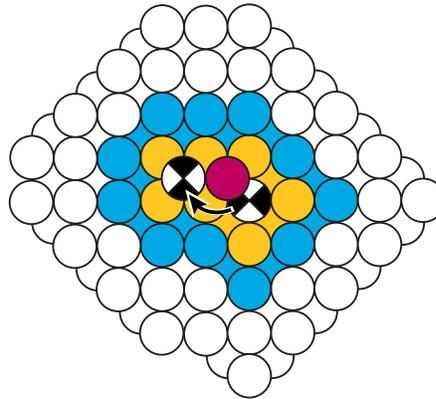

**Figure 2.** Example of a third nearest-neighbor hop: Desorption of an atom out of a perfect (001) plane. A nearest-neighbor hop to the red site is not possible.

### Diffusion of Ag monolayer islands and voids on Ag(111) and Ag(001)

Three diffusion mechanisms are possible for the diffusion of islands and voids, namely periphery diffusion, terrace diffusion, or evaporation-condensation. For the three limiting cases, the diffusion coefficient $D$ scales differently with the object radius $R$, $D \sim R^{-3}$, $R^{-2}$ and $R^{-1}$, respectively [6, 7].

The diffusion coefficients of monolayer voids and islands on Ag(111) and Ag(001) were calculated. We used a temperature of 350 K and four different radii 10, 12, 15 and 20 $r_0$, where $r_0$ denotes the nearest-neighbor distance. The initial simulation setup consists of a (111) or (001) oriented substrate of immobile atoms, respectively. On top of the substrate we place one partially filled layer of mobile atoms, forming an initially round island or void of radius $R$. During the simulation the voids and islands relax into a facetted hexagonal or rectangular equilibrium shape corresponding to the symmetry of the substrate at hand. Analysis of the simulations showed that the islands and voids diffuse almost exclusively via periphery diffusion.

$D$ can be calculated as

$$D = \frac{1}{4\Delta t}\left\langle \left|\underline{r}_C(t) - \underline{r}_C(t+\Delta t)\right|^2 \right\rangle, \tag{2}$$

where $\underline{r}_C$ denotes the void or island center. We calculated $D$ as a temporal average. Then, the diffusion coefficient can be extracted from a linear fit of $|\underline{r}_C(t) - \underline{r}_C(t+\Delta t)|^2$ as a function of the different $\Delta t$.

The diffusion coefficients and the corresponding scaling exponents are shown in figure 3. As predicted, $D$ decreases with increasing $R$. Since the (001) surface exhibits a stronger corrugation than the (111) surface, periphery diffusion processes are energetically less favorable and hence $D$ becomes generally significantly smaller. Whereas on Ag(111) the diffusion coefficient of an island is larger than the one of an equivalent void, the opposite is true on Ag(001). While not being the only important processes the following differences provide a significant contribution to these trends. On Ag(111), the detachment of the first atom out of a void edge is energetically less favorable than for an island edge, where the atom can simply diffuse around the island corner, leading to a faster island diffusion on Ag(111). On Ag(001) however, the processes are energetically very close and the diffusion along the concave island border is slower. For a more quantitative evaluation, further processes would have to be considered, like in Mehl *et al.* [8].

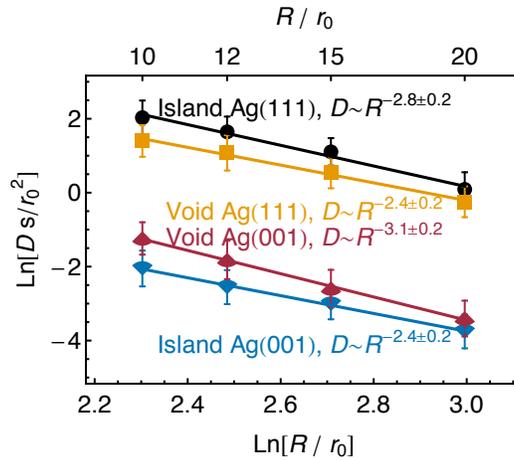

**Figure 3.** Double logarithmic plot of the monolayer island and void diffusion coefficient on Ag(111) and Ag(001) for different radii.

In spite of the observed periphery diffusion, the scaling exponents show a deviation from the predicted value minus three to larger values. Since we studied relatively small islands and voids, which exhibit a strong faceting, an influence of the void shape and the associated important single atom diffusion processes is to be expected.

**ARBITRARILY ORIENTED SURFACES**

Using one fixed surface orientation would be a rather poor approximation for most processes encountered for three-dimensional shapes, e.g. during the sintering of nanoparticles [9]. To circumvent this limitation, we developed an algorithm to determine for each appearing process the locally most suitable surface orientation. We restricted the sample of possible surface orienta-tions to the {111} and {001}-like surfaces. For the pattern recognition scheme, we mapped every {111} orientation on the (111) orientation and every {001} orientation on the (001) orientation.

To determine the matching surface orientation for a hopping process (the *process orientation*), we propose to assign each site $i$ an orientation $\underline{o}_i$. The orientation is determined by the sum of the connection vectors between the site position $\underline{r}_i$ and its occupied ($\sigma_j = 1$ else 0) neighbor site positions $\underline{r}_j$ up to the second neighbor shell (54 sites):

$$\underline{o}_i = \sum_{j=1}^{54} \sigma_j \left( \underline{r}_j - \underline{r}_i \right). \tag{3}$$

For a hopping process between two sites, the orientation of the initial plus the final site, subtracting the connection vector of the hopping atom at the final site, gives the process orientation $\underline{o}_p$:

$$\underline{o}_p = \underline{o}_i + \underline{o}_f - (\underline{r}_i - \underline{r}_f). \tag{4}$$

The term $\underline{r}_i - \underline{r}_f$ assures that the hop in the opposite direction uses the same process orientation.

The resulting process orientation is now compared with the eight possible <111> and six possible <100> surface normals. The smallest scalar product between the process orientation and a surface normal determines the best matching surface orientation. Note that this method allows a fast orientation update after a hopping process. In figure 4, the process orientation dependent surface types are shown.

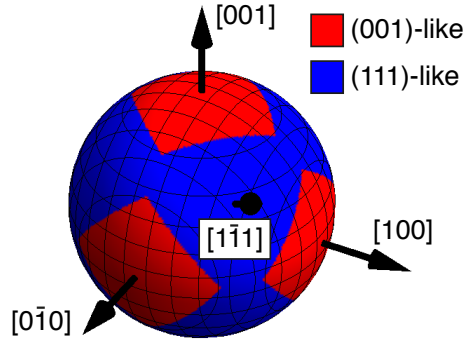

**Figure 4.** Process orientation dependent surface types. For the pattern recognition scheme, blue domains are mapped onto the (111) oriented surface; red domains are mapped onto the (001) oriented surface.

## **Particle relaxation**

As an example of a system, where multiple surface orientations appear, we simulated at $T = 350$ K the relaxation of an initially spherical particle with radius 6 $r_0$ into its facetted equilibrium shape. The particle surface consists of {111} and {001} facets, which exhibit the lowest surface energy for Ag [10]. During the simulation we measured the $T = 0$ K energy $E_0$ of the par-ticle as an approximation of the actual energy. To calculate $E_0$ we relaxed the particle to the local PES minimum. The time evolution of $E_0$ is shown in figure 5. $t = 30$ *m*s corresponds to approximately $1.1 \times 10^7$ KMC steps. As expected, $E_0(t)$ decreases with $t$.

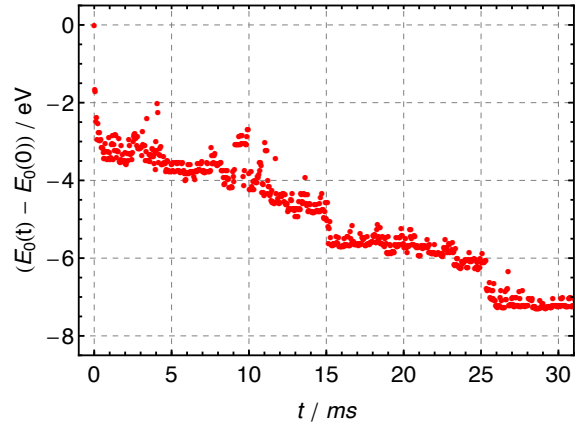

**Figure 5.** Time evolution of the $T = 0$ energy $E_0$ of an initially spherical particle with radius $6\,r_0$.

## CONCLUSIONS

In this work we presented a method, which allows for the first time SLKMC simulations of systems without a predefined global surface orientation. We propose an algorithm to determine for each hopping process a local orientation, which determines the surface orientation best suited for the description of the local environment of the process. At choice are all {111} and {001} surfaces. Moreover, the algorithm minimizes the number of processes to be stored, by utilizing the symmetry of the fcc lattice. As an application example, we simulated the relaxation of a three-dimensional spherical particle.

## ACKNOWLEDGMENTS

Financial support from the Deutsche Forschungsgemeinschaft through SFB616 'Energy Dissipation at Surfaces' is gratefully acknowledged.

## REFERENCES

1. O. Trushin, A. Karim, A. Kara and T. S. Rahman, *Phys. Rev.* B **72**, 115401 (2005).
2. A. Latz, L. Brendel and D. E. Wolf, *J. Phys.: Condens. Matter* **24**, 485005 (2012).
3. A. F. Voter, *Phys. Rev.* B **34**, 6819 (1986).
4. F. Cleri and V. Rosato, *Phys. Rev.* B **48**, 22 (1993).
5. Z.-H. Huang and R. E. Allen, *J. Vac. Sci. Technol.* A **9**, 876 (1991).
6. S. V. Khare, N. C. Bartelt, and T. L. Einstein, *Phys. Rev. Lett.* **75**, 2148 (1995).
7. K. Morgenstern *et al.*, *Phys. Rev. Lett.* **74**, 2058 (1995).
8. H. Mehl, O. Biham, O. Millo, and M. Karimi, *Phys. Rev.* B **61**, 4975 (2000).
9. F. Westerhoff, R. Zinetullin and D. E. Wolf, in *Powders and Grains*, edited by R. Garcia-Rojo, H. J. Herrmann and S. McNamara, (Balkema, Leiden, 2005), pp. 641-645.
10. Y. Wen, and J. Zhang, *Solid State Commun.* **144**, 163 (2007).